\documentclass[aps,prl,twocolumn,floatfix,superscriptaddress,showpacs,longbibliography]{revtex4-1}

\usepackage{amsfonts}
\usepackage{amsmath}
\usepackage{amssymb}
\usepackage{mathtools}
\usepackage{graphicx}
\usepackage{color}
\usepackage{bbm}
\usepackage[hidelinks]{hyperref}
\hypersetup{
     colorlinks   = true,
     citecolor    = blue,
     linkcolor    = blue
}
\usepackage{subfigure}
\usepackage{mathptmx}
\usepackage{times,txfonts}
\usepackage{siunitx}

\newcommand{\tr}{{\rm Tr}}
\newcommand{\E}{{\mathcal E}}
\newcommand{\A}{{\mathcal A}}
\newcommand{\U}{{\mathcal U}}
\newcommand{\der}[2]{\frac{\partial #1}{\partial #2}}
\newcommand{\dsec}[2]{\frac{{\partial}^2 #1}{\partial {#2}^2}}
\newcommand{\pton}[1]{\left(#1\right)}
\newcommand{\nobibentry}[1]{{\let\nocite\ignore\bibentry{#1}}}
\newcommand{\re}[1]{\text{Re}\,#1}
\newcommand{\im}[1]{\text{Im}\,#1}
\newcommand{\bibfnamefont}[1]{#1}
\newcommand{\bibnamefont}[1]{#1}
\newcommand{\average}[1]{\left<#1\right>}
\newcommand{\ket}[1]{\left\vert#1\right\rangle}
\newcommand{\bra}[1]{\left\langle#1\right\vert}
\newcommand{\MAGM}[1]{{\color{green} #1}}
\newcommand{\moha}[1]{{\color{red} #1}}

\begin{document}

\title{Using polarons for sub-\si{\nano\kelvin} quantum non-demolition thermometry in a Bose-Einstein condensate}

\author{Mohammad~Mehboudi} 
\affiliation{ICFO---Institut de Ci\`encies Fot\`oniques, The Barcelona Institute of Science and Technology, 08860 Castelldefels (Barcelona), Spain}
\affiliation{Departament de F\'isica, Universitat Aut\`onoma de Barcelona, E-08193 Bellaterra, Spain}
\author{Aniello~Lampo} 
\affiliation{ICFO---Institut de Ci\`encies Fot\`oniques, The Barcelona Institute of Science and Technology, 08860 Castelldefels (Barcelona), Spain}
\author{Christos Charalambous} 
\affiliation{ICFO---Institut de Ci\`encies Fot\`oniques, The Barcelona Institute of Science and Technology, 08860 Castelldefels (Barcelona), Spain}
\author{Luis~A.~Correa}
\affiliation{School of Mathematical Sciences and Centre for the Mathematics and Theoretical Physics of Quantum Non-Equilibrium Systems,
	The University of Nottingham, University Park, Nottingham NG7 2RD, United Kingdom}
\affiliation{Kavli Institute for Theoretical Physics University of California, Santa Barbara, CA 93106}
\author{Miguel~\'Angel~Garc\'ia-March} 
\affiliation{ICFO---Institut de Ci\`encies Fot\`oniques, The Barcelona Institute of Science and Technology, 08860 Castelldefels (Barcelona), Spain}
\author{Maciej~Lewenstein}
\affiliation{ICFO---Institut de Ci\`encies Fot\`oniques, The Barcelona Institute of Science and Technology, 08860 Castelldefels (Barcelona), Spain}
\affiliation{ICREA, Llu\'is Companys 23, E-08010 Barcelona, Spain}
        
\begin{abstract}
We introduce a novel minimally-disturbing method for sub-\si{\nano\kelvin} thermometry in a Bose-Einstein condensate (BEC). Our technique is based on the Bose-polaron model; namely, an impurity embedded in the BEC acts as the thermometer. We propose to detect temperature fluctuations from measurements of the position and momentum of the impurity. Crucially, these cause minimal back-action on the BEC and hence, realize a non-demolition temperature measurement. Following the paradigm of the emerging field of \textit{quantum thermometry}, we combine tools from quantum parameter estimation and the theory of open quantum systems to solve the problem in full generality. We thus avoid \textit{any} simplification, such as demanding thermalization of the impurity atoms, or imposing weak dissipative interactions with the BEC. Our method is illustrated with realistic experimental parameters common in many labs, thus showing that it can compete with state-of-the-art \textit{destructive} techniques, even when the estimates are built from the outcomes of accessible (sub-optimal) quadrature measurements.
\end{abstract}

\date{\today}
\maketitle

\noindent\textit{Introduction---} 
The ongoing efforts in the development of quantum technologies is strongly fuelled by their many anticipated practical applications \cite{2017Celi}. In the process, we are already benefiting from striking experimental advances and much deeper theoretical insights. In particular, ultracold atomic gases are a key platform for quantum technologies due to their potential for quantum simulation \cite{2008Bloch,2012Lewenstein}. Nonetheless, operating a quantum simulator requires very precise tuning of the parameters of the experiment, so as to ensure that the simulated system behaves as intended. In particular, a precise temperature control is essential, for instance, for the reconstruction of the equation of state of the system \cite{2012Bloch}. 

In current experimental setups, the main thermometric techniques are based on time-of-flight measurements either directly on the BEC \cite{leanhardt2003cooling,gati2006noise,gati2006primary}, or on impurities embedded in it \cite{olf2015thermometry,spiegelhalder2009collisional}. In the former case, temperatures of few \si{\nano\kelvin}, or even sub-\si{\nano\kelvin} might be estimated efficiently, although at the price of destroying the BEC. On the contrary, the latter protocols are less destructive, albeit efficient at relatively ``large'' temperatures of $ \sim 100 $ \si{\nano\kelvin}. Interestingly, recent proposals have discussed minimally disturbing interferometric setups in which the temperature is mapped onto a relative phase on a probe \cite{stace2010limits,martin2013berry,sabin2014impurities}, however, the underlying models are very simple. 

An effective non-demolition thermometric technique in the sub-\si{\nano\kelvin} regime is thus still missing. Any such strategy should be build upon a {\it comprehensive theoretical description} and be capable of informing the choice of the most sensitive temperature-dependent quantities to be measured. Here, we propose what is, to the best of our knowledge, the first experimentally feasible quantum non-demolition technique to measure the temperature of a BEC in the sub-\si{\nano\kelvin} domain. It is based on the Bose polaron problem, i.e., interrogation of an impurity that is embedded in the condensate, while causing minimal disturbance to the cold atomic gas. The impurity problem has been intensively studied in the context of polaron physics in strongly-interacting Fermi~\cite{Schirotzek2009, Kohstall2012, Koschorreck2012,
	MassignanPolRev2014, PhysRevLett.116.105302,Lan2014, Levinsen2014, Schmidt2012,PhysRevA.97.063619} or Bose gases \cite{PhysRevA.88.053632,Cote2002, Massignan2005, Cucchietti2006, 
	Lampo2017, Pastukhov2017, Yoshida2017, Guenther2018, Lampo2018},
as well as in solid state physics \cite{Landau48, Devreese2009, Alexandrov2009}, and mathematical physics \cite{Lieb1958, Lieb1997, Frank2010, Anapolitanos2013, Lim2018}. We specifically \textit{avoid} any unjustified simplifications---such as complete thermalization of the impurities at the BEC temperature---and investigate the problem in its full generality. The usefulness of our proposed technique is finally illustrated with typical experimental parameters.

In our analysis, we benefit from the toolbox of the emergent field of \textit{quantum thermometry} \cite{mehboudi2018thermometry}, which combines quantum estimation theory and the theory of open quantum systems. This will allow us to compare the \textit{ultimate} precision bounds on temperature estimation with the thermal sensitivity of concrete experimentally feasible measurements.
\newline

\noindent\textit{The model---} 
Let us consider an impurity of mass $ m_{\rm I} $ (acting as the temperature \textit{probe}) embedded in a BEC of atoms of mass $ m_{\rm B} $, chemical potential $ \mu $, and interatomic coupling strength $ g_{\rm B} $. The condensate (which makes up the \textit{sample}), is confined in a one-dimensional harmonic well of frequency $ \omega_{\rm B} $, leading to a parabolic Thomas-Fermi potential with radius $ R = \sqrt{2\mu/m_{\rm B}\omega^2_{\rm B}} $. In turn, the impurity is trapped in a harmonic potential of frequency $ \Omega $. Finally, the interspecies coupling (i.e., the probe-sample interaction) is denoted by $ g_{\rm IB} $. Our aim is to estimate the temperature $ T $ of the BEC as precisely as possible, while diminishing the ensuing disturbance. 

We note that the global probe-sample Hamiltonian can be thought-of as a quantum Brownian motion model consisting of the following contributions:
\begin{align}
\hat{H} &= \hat{H}_{\rm I}+\hat{H}_{\rm B}+\hat{H}_{\rm int} \nonumber\\
		&= \frac{{\hat{p}}^2}{2m_{\rm I}}+\frac{m_{\rm I}\Omega^2}{2}\hat{x}^2 + \sum\nolimits_kE_k\hat{b}^{\dagger}_k\hat{b}_k + \sum\nolimits_k\hbar g_k\,\hat{x}\left(\hat{b}_k+\hat{b}^\dagger_k\right).
\label{eq:hamiltonian}
\end{align}
Here, $ \hat{H}_{\rm I} $ stands for the free Hamiltonian of the impurity, while the term $ \hat{H}_{\rm B} $ represents the BEC and encompasses all the interacting modes of the atomic gas. Finally, $\hat{H}_{\rm int}$ is the interaction between the impurity and the gas.    
In the second line of Eq.~\eqref{eq:hamiltonian}, however, we express the BEC degrees of freedom in terms of the operators $ \hat{b}_k $ and $ \hat{b}_k^{\dagger} $, that stand for the annihilation and creation operators of a Bogoliubov mode with energy $ E_k $. 
This new representation is nothing but the result of a standard Bogoliubov transformation, which diagonalizes $ \hat{H}_{\rm B} $ and maps it into a non-interacting form. 
In this picture, the last term in the second line of \eqref{eq:hamiltonian} accounts for the interactions between the impurity and the Bogoliubov modes. 
The interaction strength between the $k$th Bogoliubov mode and the impurity is given by $ g_k $, (see \cite{Lampo2018} for details). 
Note that such interactions exhibit, in general, a non-linear dependence on the position of the impurity. The linear form presented in Eq.~\eqref{eq:hamiltonian} is valid only near the center of the confining potential of the BEC, i.e., when $ x/R \ll 1 $. Of course, this leads to constraints on the values of the system parameters that have been discussed in \cite{Lampo2018}. We underline that the values of the physical quantities considered in the following fulfill the conditions associated to the linear approximation in the interaction Hamiltonian.

In general, the spectrum of a Hamiltonian like $ \hat{H} $ may not be bounded from below \cite{Caldeira1983a,Caldeira1983b,weiss2008quantum}. To rule out this eventuality, it is common practice to \textit{shift} the frequency of the Brownian particle as $ \Omega^2 \mapsto \tilde{\Omega}^2 \coloneqq \Omega^2 + 2 \sum_k g_k^2/ \omega_k^2 $ to compensate for the distortion caused by the coupling to the bath \cite{PhysRevA.96.062103,hovhannisyan2017probing}. Here, however, we will avoid adding any terms ``by hand'', since Eq.~\eqref{eq:hamiltonian} has been consistently derived from first principles~\cite{Lampo2018}. Instead, we will limit ourselves to choose parameters which fulfill the conditions described in~\cite{Lampo2018} that guarantee the positivity of the spectra.

Treating the impurity as a Brownian particle coupled to a bath of Bogoliubov modes, allows us to exploit well-established techniques from the theory of open quantum systems. Specifically, the motion of the probe is described \textit{exactly} by the second-order differential equation \cite{weiss2008quantum}
\begin{equation}\label{EqDiffFin}
\frac{d^2\hat{x}(t)}{dt^2} + \Omega^2 \hat{x}(t)+\frac{d}{dt}\int^{t}_0 ds\,\Gamma(t-s)\,\hat{x}(s)=\frac{\hat{B}(t)}{m_{\rm I}}.
\end{equation}
This is the quantum counterpart of the Langevin equation, introduced in 1909 for (classical) Brownian motion \cite{Langevin1908, MazoBook}.
The term $\hat{B}(t)$ on the right-hand side of Eq.~\eqref{EqDiffFin} reads 
\begin{equation}
\hat{B}(t) \coloneqq \sum_k \hbar g_{k}\left(\hat{b}^{\dagger}_k e^{-i\omega_k t}+\hat{b}_k e^{i\omega_k t}\right),
\end{equation}
and plays the role of a stochastic driving force. Memory effects enter in the dynamics through the \textit{damping kernel} 
\begin{equation}
\label{DampingKernel}
\Gamma(\tau) \coloneqq \frac{1}{m_{\rm I}}\int^{\infty}_0 d\omega\,\frac{J(\omega)}{\omega}\cos(\omega\tau),
\end{equation}
where $ J(\omega) \coloneqq \sum_{k\neq 0} \hbar g_k^2 \delta(\omega-\omega_k) $ is the \textit{spectral density}. For our model, this is given by
\begin{align}
\label{eq:spectra_dens}
J(\omega) = m_{\rm I}\gamma\,\frac{\omega^4}{\omega^3_{\rm B}}\,\Theta\left(\omega-\omega_{\rm B}\right),
\end{align}
where $ \gamma \coloneqq (2g_{\rm B}/m_{\rm I}\omega_{\rm B}R^3)\times(\eta\,\mu/\hbar\omega_{\rm B})^2 $ and $ \eta \coloneqq g_{\rm IB}/g_{\rm B} $ \cite{Lampo2018}. The Heaviside function $ \Theta(\cdot) $ introduces an ultraviolet cutoff, which regularizes the diverging high-frequency behavior. Importantly, the long-time dynamics of the impurity---our main focus---does not depend on the details of the cutoff \cite{Lampo2017} but rather, on the low frequency behaviour of $ J(\omega) $. This is dictated by the exponent of $ \omega $ in the pre-factor of Eq.~\eqref{eq:spectra_dens}---i.e., the ``Ohmicity'' parameter \cite{PhysRevA.96.062103}.

Eq.~\eqref{EqDiffFin} can be solved via Laplace transform (that is, $ \mathcal{L}_z[f(t)] \coloneqq \int_0^\infty dt\,e^{-t z} f(t) $). In particular, the steady-state variances in position and momentum are given by
\begin{subequations}
\begin{align}
&\langle\hat{x}^2\rangle = \frac{\hbar}{2\pi}\int^{+\omega_{\rm B}}_{-\omega_{\rm B}}d\omega\,\coth{(\hbar\omega/2k_{\rm B}T)}\,\tilde{\chi}''(\omega),\label{X2Trap2}\\
&\langle\hat{p}^2\rangle = \frac{\hbar m^2_{\rm I}}{2\pi}\int^{+\omega_{\rm B}}_{-\omega_{\rm B}}d\omega\,\omega^2\coth{(\hbar\omega/2k_{\rm B}T)}\,\tilde{\chi}''(\omega),
\label{P2Trap2}
\end{align}
\label{eq:covariances}
\end{subequations}
where $ \langle\cdots\rangle $ denotes steady-state averaging and $ \tilde{\chi}''(\omega) $ stands for the \textit{response function} that reads as
\begin{align}\label{ResponseFunction}
\tilde{\chi}''(\omega) = \frac{1}{m_{\rm I}}\frac{\omega\,\zeta(\omega)}{\left[\omega\,\zeta(\omega)\right]^2+
	\left[\Omega^2-\omega^2+\omega\,\theta(\omega)\right]^2}.
\end{align}
$ \zeta\left(\omega\right) $ and $ \theta\left(\omega\right) $ in Eq.~\eqref{ResponseFunction} are, respectively, the real and imaginary parts of $ \mathcal{L}_{\tilde{z}}\left[\Gamma(t)\right] $ evaluated at $ \tilde{z} = -i\omega+0^{+} $. Eqs.~\eqref{DampingKernel}--\eqref{ResponseFunction} thus allow to determine the steady-state covariances of the impurity as a function of the system parameters---in particular, the temperature of the BEC and the dissipation strength $ \gamma $ \cite{PhysRevA.96.062103}. Note that, since the Hamiltonian \eqref{eq:hamiltonian} is \textit{linear} in the quadratures, Eqs.~\eqref{eq:covariances} fully characterize the steady state of the impurity (together with $ \langle\hat{x}\rangle = \langle\hat{p}\rangle = 0 $). 
\newline

\noindent\textit{Thermometric performance---}
The inherent errors from quantum measurements give rise to statistical uncertainty on the temperature estimate. Quantum estimation theory allows us to place fundamental limits on the ``error bars'' of the final temperature reading, and even to rank the various temperature-dependent properties of the probe according to their \textit{thermal sensitivity}. For instance, let us build our temperature estimate from a large set of $ \nu $ outcomes of independent measurements of some impurity observable $ \hat{O} $ \footnote{Note that we are not only assuming to work with a large dataset, but also that the estimator which maps measurement outcomes to temperature estimates is \textit{unbiased} \cite{braunstein1994statistical}.}. We stress that these are either measurements performed on \textit{independent} impurity atoms, or measurements on the same probe, but paced so that the BEC-impurity composite has time to reset to its stationary state every time. By mere propagation of errors, the uncertainty of the temperature inferred from such dataset would read \cite{toth2014quantum, braunstein1994statistical}
\begin{align}\label{eq-error-propagation}
\delta T({\hat O}) \coloneqq \frac{\Delta {\hat O}}{\sqrt{\nu~\chi_T^2({\hat{O}})}},
\end{align}
where $\Delta^2{\hat O} \coloneqq \langle{\hat O}^2\rangle - \langle\hat O\rangle^2 $ stands for the variance of $ \hat{O} $ calculated on the stationary marginal of the impurity $ \hat{\varrho}_{\rm I}(T) $, and $ \chi_T({\hat{O}}) \coloneqq \partial_{\xi}\,{\rm tr}\,[\hat{\varrho}_{\rm I}(\xi)~\hat O]_{\xi = T}$ represents its (static) temperature susceptibility.

In order to assess the performance of $ \hat{O} $, it is essential to know which is the \textit{minimum possible} uncertainty (i.e., $ (\delta T)_{\min}\coloneqq\inf_{\hat{O}}\delta T(\hat{O}) $). To this end, we introduce the symmetric logarithmic derivative (SLD) $ \hat\Lambda_T $, implicitly defined as
\begin{align}\label{eq-SLD}
	{\hat \Lambda}_T\,\hat \varrho_{\rm I}(T) + \hat \varrho_{\rm I}(T)\,{\hat \Lambda}_T \equiv 2 \left.\partial_{\xi}\,{\hat\varrho_{\rm I}(\xi)}\,\right\vert_{\xi = T}.
\end{align}
Coming back to the definition of $ \chi_T(\hat{O}) $, we notice that $\chi_T({\hat{O}}) = \frac{1}{2}\langle \hat O\,{\hat \Lambda}_T + {\hat \Lambda}_T \,{\hat O}\rangle - \langle\hat{O}\rangle\langle\hat{\Lambda}_T\rangle $, while $ \chi_T(\hat{\Lambda}_T) = \Delta^2\hat{\Lambda}_T $. Making use of the fact that $ \Delta\hat{O}\,\Delta\hat{\Lambda}_T \geq \chi_T(\hat{O}) $ allows to turn Eq.~\eqref{eq-error-propagation} into
\begin{align}\label{eq-QCRB}
\delta T({\hat O}) \geq \frac{1}{\sqrt{\nu~\Delta^2{\hat \Lambda}_T}} \coloneqq \frac{1}{\sqrt{\nu~{\cal F}(T)}},
\end{align}
where we have introduced the quantum Fisher information (QFI) \ ${\cal F}(T) \coloneqq \Delta^2{\hat \Lambda}_T $. Eq.~\eqref{eq-QCRB} is nothing but the quantum Cram\'er-Rao bound \cite{braunstein1994statistical}, and sets the ultimate lower limit on the statistical error. Furthermore, by simply replacing $ {\hat O} $ by $ {\hat \Lambda_T} $ in \eqref{eq-error-propagation}, we can see that this bound is \textit{saturated} by performing complete projective measurements onto the eigenbasis of the SLD. 
\newline

\noindent\textit{Results---} Owing to the simplified Hamiltonian in Eq.~\eqref{eq:hamiltonian} we can write the SLD and the QFI for temperature estimation solely in terms of the variances in Eqs.~\eqref{eq:covariances} \cite{monras2013phase,PhysRevA.96.062103}, i.e.,
\begin{subequations}
\begin{align}
{\hat \Lambda}_T &= C_x \left({\hat x}^2 - \langle{\hat x}^2\rangle\right) + C_p \left({\hat p}^2 - \langle{\hat p}^2\rangle\right),\label{eq-SLD-Gaussian}\\
{\cal F}(T) &= 2\,C_x^2\,\langle{\hat x}^2\rangle^2 + 2\,C_p^2\,\langle{\hat p}^2\rangle^2 - \hbar^2\, C_x\,C_p \label{eq-QFI-COV}, 
\end{align}
\label{eq:SLD_QFI}
\end{subequations}
where the coefficient $ C_x $ is given by
\begin{align}
	C_x &= \frac{4\,\langle{\hat p}^2\rangle^2\,\chi_T({\hat x}^2) + \hbar^2\,\chi_T({\hat p}^2)}{8\,\langle{\hat x}^2\rangle^2\,\langle{\hat p}^2\rangle^2 - \hbar^4/2},
\end{align}
and $ C_p $ can be obtained by simply exchanging $ \hat{x} $ and $ \hat{p} $. That is, by repeatedly measuring the observable \eqref{eq-SLD-Gaussian} on the impurity, the temperature of the BEC can be estimated with the minimum possible uncertainty. We are now in the position to plug in realistic numbers into the exact steady-state marginal for the probe and explore the thermal sensitivity of our non-demolition thermometric protocol at ultra-low temperatures.

\begin{figure}
	\includegraphics[width=\linewidth]{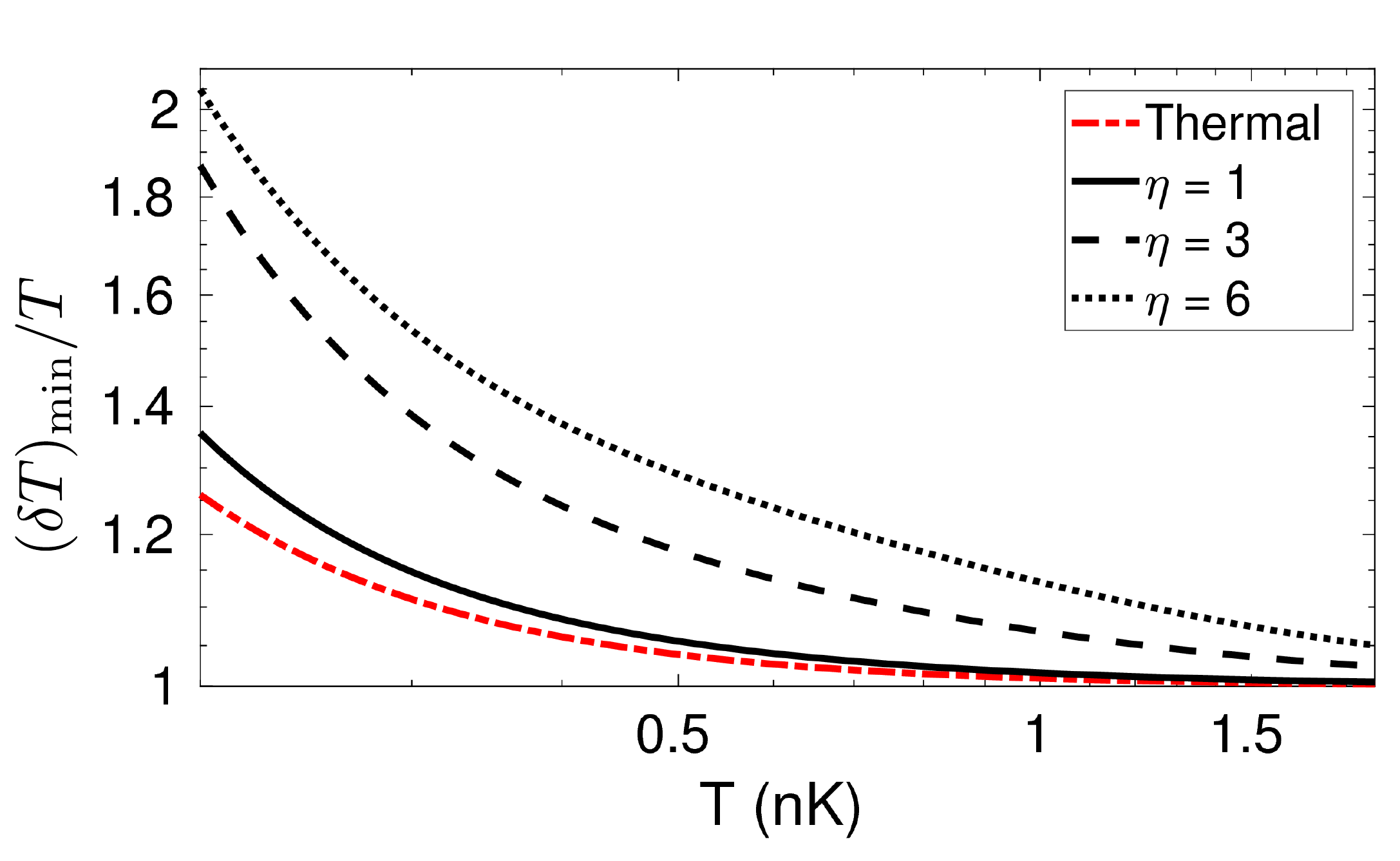}
	\caption{(color online) (black) Optimal relative error $ (\delta T)_{\rm min}/T $ ($\nu = 1$) as a function of the temperature of the BEC in a logarithmic scale. Specifically, we work with impurities of Yb in a sea of ultra-cold K. The temperature range for the BEC is $ 200\,\si{\pico\kelvin} \leq T \leq 2\,\si{\nano\kelvin} $. The trapping frequency of the gas (with $ N = 5000 $ atoms) was set to $\omega_{\rm B} = 2\pi\times100\,\si{\hertz}$, while $ \Omega = 2\pi\times10\,\si{\hertz} $ ($ g_{rm B} = 3\times10^{-39}\,\si{\joule\meter} $). Different probe-sample coupling ratios $ \eta = g_{\rm IB}/g_{\rm B} $ were considered, namely (solid) $ \eta = 1 $, (dashed) $ \eta = 3 $, and (dotted) $ \eta = 6 $. For comparison, we also depicted the relative error of a fully thermalized impurity (i.e., $ \eta \rightarrow 0 $) (dot-dashed red). Note that, for $ \eta = 1 $, the relative error can be kept below $ \sim 14 $\% from only $ \nu \sim 100 $ measurements. This is quantitatively close to state-of-the-art \textit{destructive} experimental techniques. See text for discussion.}
	\label{fig-QFI}
\end{figure}

\begin{figure*}
	\begin{tabular}{c}
		\includegraphics[width=\columnwidth]{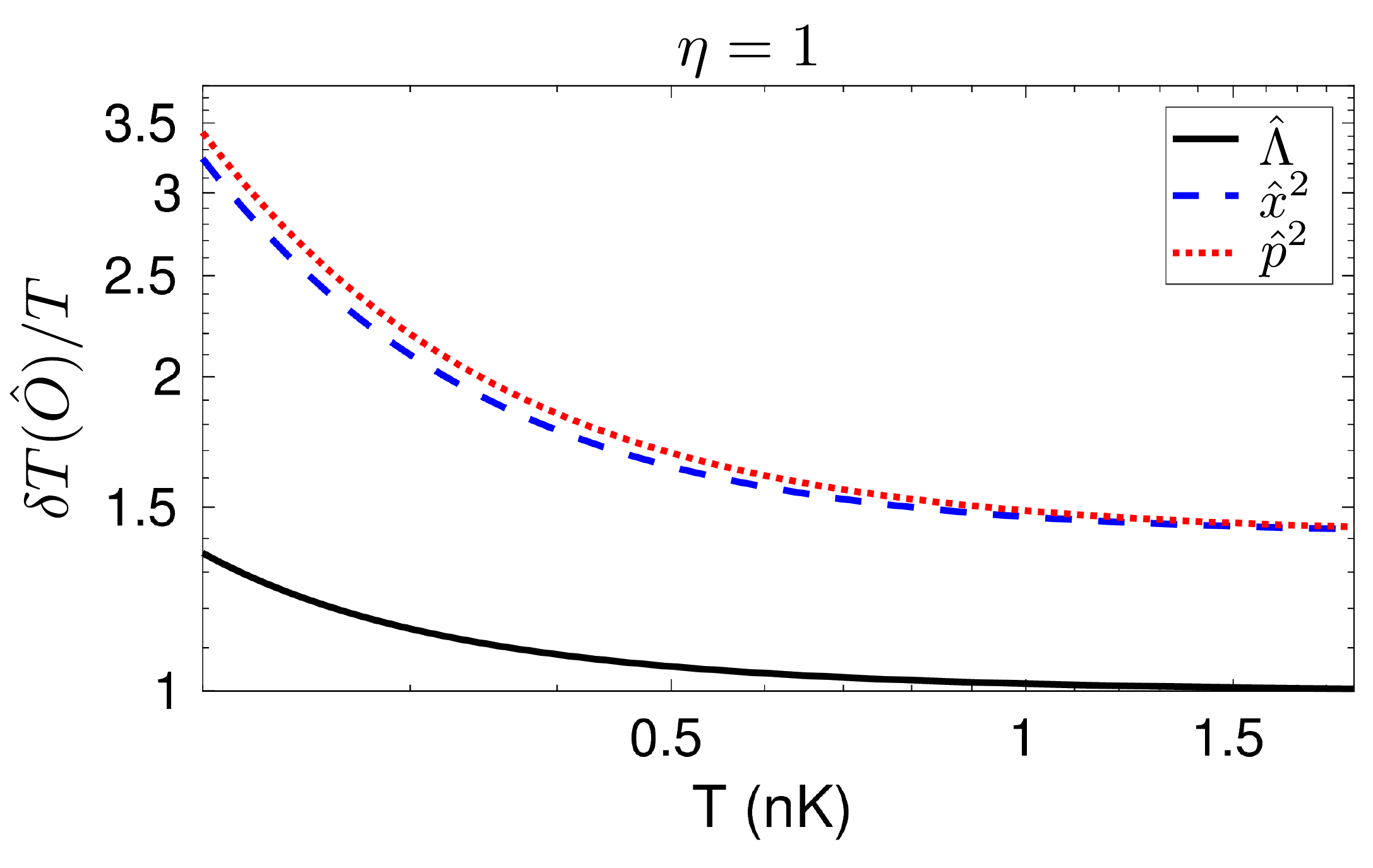}
		\includegraphics[width=\columnwidth]{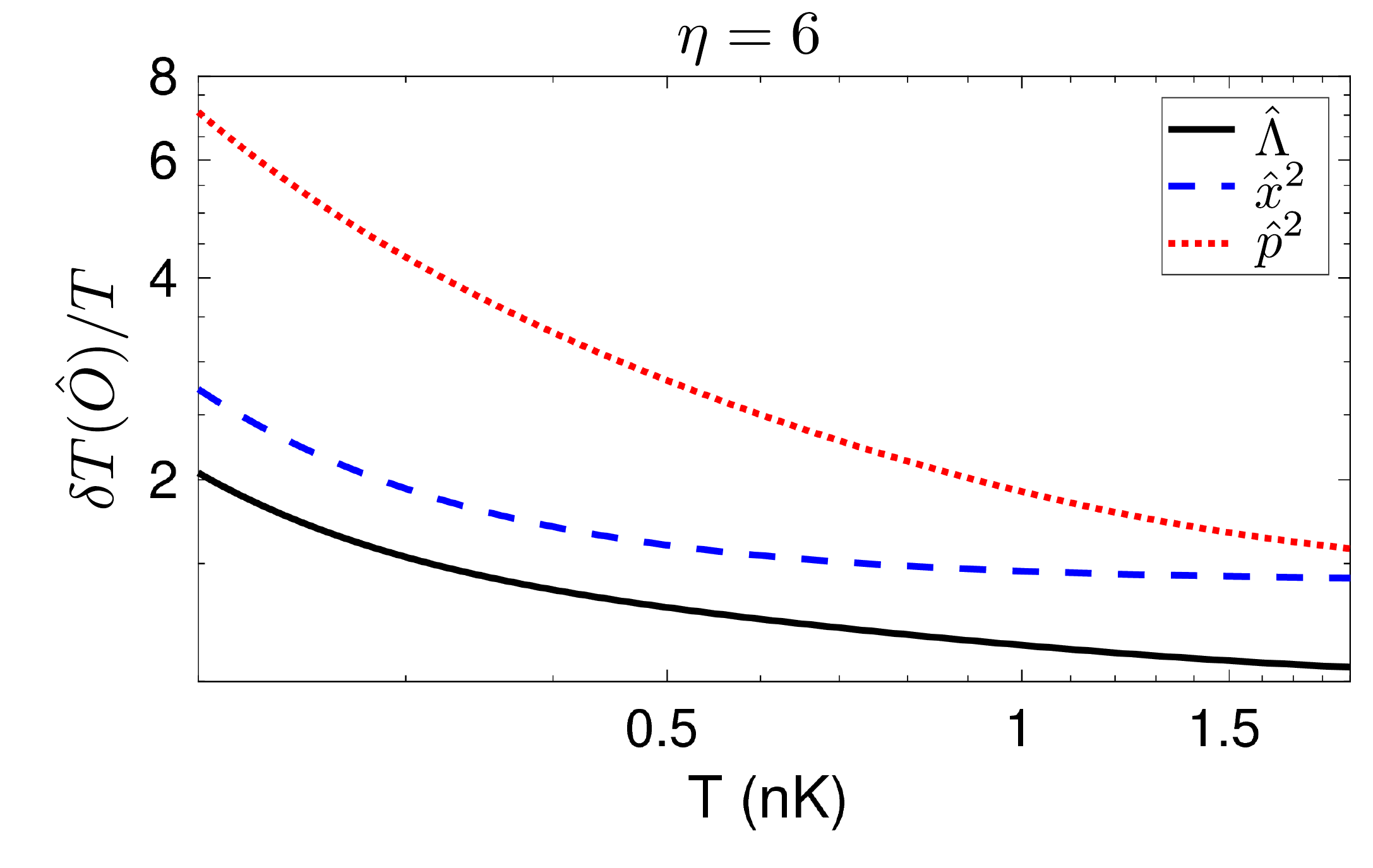} 
	\end{tabular}
	\caption{(color online) (dashed blue) Relative error for the position quadrature $ \delta T(\hat{x}^2)/T $ and (dotted red) the momentum quadrature $ \delta T(\hat{x}^2)/T $ as a function of $ T $ for (left panel) $ \eta = 1 $ and (right panel) $ \eta = 6 $. All parameters are the same as in Fig.~\ref{fig-QFI}. The minimum relative error (solid black) is superimposed for reference. Even though both measurement schemes are sub-optimal, they still might allow to draw estimates with relative errors as low as $ 18\% $ for $ \nu \sim 400 $. Note that for $T\geq 0.5$~nK and by using the same data size $\nu=400$, one can achieve a relative error below $10\%$.}
	\label{fig-Sub-Optimal}
\end{figure*}

As an illustrative example, we will work with a BEC of K atoms containing Yb impurities. The qualitative picture would remain essentially unaltered regardless of the atomic species considered. In Fig.~\ref{fig-QFI}, we plot the optimal relative error $ \sqrt{\nu}\,(\delta T)_{\min}/T = \big(T \sqrt{\mathcal{F}_T}\big)^{-1} $ for various probe-sample coupling strengths and temperatures ranging from $ 200\,\si{\pico\kelvin} $ to $ 2\,\si{\nano\kelvin} $. Specifically, keeping the interatomic and interspecies couplings comparable (i.e., $ \eta = 1 $) would allow to achieve a relative error below $ 14 $\% from as few as $ 100 $ measurement outcomes. That is, polaron thermometry outperforms the interferometric technique proposed in Ref.~\cite{sabin2014impurities} by an order of magnitude. More importantly, unlike state-of-the-art experimental methods (e.g., \cite{leanhardt2003cooling,gati2006primary,gati2006noise,sherson2010single,olf2015thermometry}), ours is \textit{non-destructive}.

We note, however, that the \textit{stronger} the probe sample interaction, the \textit{worse} the estimation. Likewise, it can be clearly seen that, for strong dissipation, the impurity deviates significantly from a thermal state at the temperature of the sample. The first observation seems to be in striking contradiction with the main results of \cite{PhysRevA.96.062103}, where a substantial dissipation-driven \textit{enhancement} was reported at low temperatures. Note however, that the temperature range considered in Fig.~\ref{fig-QFI} \textit{does not} qualify as ``low'', according to the criteria of Ref.~\cite{PhysRevA.96.062103}, namely $ T \ll \hbar\omega_{\rm B}/k_B $ (here, $ T \sim \hbar\omega_{\rm B}/k_B $). When it comes to the second observation, it is worth highlighting that the divergence between the exact steady state of the impurity and a fully thermalised probe can be sizeable in the $ \si{\pico\kelvin} $ range. This only comes to reinforce the idea that simple dissipation models, such as a Gorini-Kossakowski-Lindblad-Sudarshan master equation \cite{lindblad1976generators,gorini1976completely} are not suitable for this type of analysis.

Recall that the above discussion assumes that the optimal measurement of Eq.~\eqref{eq-SLD-Gaussian} can be implemented. In practice, however, such a mixture of covariances with temperature-dependent coefficients might be difficult to realize; the \textit{bare} quadratures $ \langle \hat{x}^2 \rangle $ or $ \langle\hat{p}^2\rangle $ being easier to measure. The relative error of estimates based on these is benchmarked against the ultimate lower bound in Figs.~\ref{fig-Sub-Optimal}. Note that, at $ \eta = 1 $, $ \langle\hat{x}^2\rangle $ and $ \langle\hat{p}^2\rangle $ perform similarly, while at stronger coupling, the position quadrature becomes a significantly better temperature estimator. Also, under stronger dissipation, $ \langle \hat{x}^2 \rangle $ gets closer to the optimal setting. Importantly, our approach remains practically useful regardless of the strict sub-optimality of $ \hat{x}^2 $---temperature estimates with $ \delta T(\hat{x}^2)/T < 18\% $ (or in the domain $T\geq 0.5$~nK, with $ \delta T(\hat{x}^2)/T < 10\% $) can still be constructed from relatively small datasets of $ \nu \sim 400 $.
\newline

\noindent\textit{Conclusions---}
We have shown that impurities immersed in a BEC can be exploited as temperature sensors. The key features of such thermometric scheme are that (i) the temperature is estimated by monitoring the impurity atoms \textit{only}---the BEC itself does not need to be measured destructively, (ii) it can compete with state-of-the-art thermometric techniques in the sub-\si{\nano\kelvin} range, and (iii) the underlying analysis does not assume thermalization of the impurity at the temperature of the BEC, but rather takes fully into account the strong correlations built up between probe and sample. 

In particular, we considered a cold atomic gas and an impurity both harmonically confined in 1D at different trapping frequencies. Assuming that the impurity remains localised around the minimum of the potential, allowed us to ``linearize'' the model. We obtained the exact stationary state of the impurity from the corresponding quantum Langevin equation and, using standard tools from quantum estimation theory, we could eventually calculate the minimum possible statistical uncertainty for a temperature measurement. In particular, owing to our analysis being exact, we could verify that the usual assumption of full thermalization for the impurity at the temperature of the sample overestimates the performance of the scheme for typical parameters in the \si{\pico\kelvin}--\si{\nano\kelvin} range.

We showed that, with only $100$ measurements, the relative error can be kept below $ 14 \% $ for temperatures as low as $ 200\,\si{\pico\kelvin} $. Importantly, we could also show that feasible sub-optimal quadrature measurements---specifically, $ \hat{x}^2 $---allow for similar performances with limited resources (i.e., datasets of just few hundreds of independent measurements). Interestingly, we found that increasing the probe-sample coupling does not improve the sensitivity of the protocol in the temperature range under study due to the comparatively low typical trapping frequencies ($ 60 $--$ 70 $ \si{\hertz}).

Even though we illustrate our results with Yb impurities in a cold gas of K atoms, our approach is completely general and could be straightforwardly applied without limitations to other atomic species and temperature ranges. Similar results are also expected in the 2D and 3D cases. In particular, such an extension is straightforward for \textit{homogeneous} BECs, the same position squeezing effects giving rise to the enhanced sensitivity of $ \hat{x}^2 $ are known to occur \cite{Lampo2017}; the main difference would be a larger Ohmicity in Eq.~\eqref{eq:spectra_dens}.

In order to bring these promising quantum non-demolition thermometers a step closer to experimental demonstrations, it would be interesting to study how the unavoidable non-linearities could affect our results. Exploring whether the entanglement between two impurities embedded in the BEC---recently studied in~\cite{charalambous2018two}---can be used to boost thermometric performance also remains an open challenge. 
\newline

\begin{acknowledgements} 
We acknowledge fruitful discussions with A. Sanpera and A. Ac\'in. 
This work has been funded by the Spanish Ministry MINECO (project QIBEQI FIS2016-80773-P, National Plan 15 Grant: FISICATEAMO No. FIS2016-79508-P, SEVERO OCHOA No. SEV-2015-0522, FPI), European Social Fund, Fundaci\'o Privada Cellex, Generalitat de Catalunya (AGAUR Grant No. 2017 SGR 1341, SGR 1381 and CERCA/Program), ERC AdG OSYRIS, EU FETPRO QUIC, the National Science Centre, Poland-Symfonia Grant No. 2016/20/W/ST4/00314, ERC StG GQCOP No. 637352, the COST Action MP1403: ``Nanoscale quantum optics'', and US National Science Foundation under Grant No. NSF PHY-1748958. Financial support through a scholarship from Programa M\`{a}sters d'Excel\textperiodcentered l\`{e}ncia (Fundaci\'{o} Catalunya-La Pedrera) is appreciated. 
LAC thanks the Kavli Institute for Theoretical Physics for their warm hospitality during the program ``Thermodynamics of quantum systems: Measurement, engines, and control''.
\end{acknowledgements}

\bibliographystyle{apsrev4-1}
\bibliography{Refs}

\end{document}